\documentclass[aps,showpacs,pre,superscriptaddress,twocolumn]{revtex4}
\usepackage{epsfig}
\newcommand{\be}{\begin{equation}}
\newcommand{\ee}{\end{equation}}
\newcommand{\bea}{\begin{eqnarray}}
\newcommand{\eea}{\end{eqnarray}}
\newcommand{\sn}{{\rm sn}}
\newcommand{\ds}{{\rm ds}}
\newcommand{\cs}{{\rm cs}}
\newcommand{\ns}{{\rm ns}}
\newcommand{\dn}{{\rm dn}}
\newcommand{\cn}{{\rm cn}}
\newcommand{\sech}{{\rm sech}}
\begin{document}
\title{Discrete Nonlinear
Schr{\"o}dinger Equations with arbitrarily high order
nonlinearities}
\author{Avinash Khare}
\affiliation{Institute of Physics, Bhubaneswar, Orissa 751005, India}
\author{Kim~{\O}. Rasmussen}
\affiliation{Theoretical Division and Center for Nonlinear Studies,
Los Alamos National Laboratory, Los Alamos, New Mexico, 87545, USA}
\author{Mario Salerno}
\affiliation{Dipartimento di Fisica \textquotedblleft E.R.
Caianiello", \\Istituto Nazionale di Fisica Nucleare (INFN)
Sezione di Napoli-Gruppo Collegato di Salerno, Consorzio Nazionale
Interuniversitario per le Scienze Fisiche della Materia (CNISM),
Universit\'{a} di Salerno, I-84081 Baronissi (SA), Italy}
\author{Mogens R. Samuelsen}
\affiliation{Department of Physics, The Technical University of Denmark,
DK-2800 Kgs. Lyngby, Denmark}
\author{Avadh Saxena}
\affiliation{Theoretical Division and Center for Nonlinear Studies,
Los Alamos National Laboratory, Los Alamos, New Mexico, 87545, USA}
\date{\today}


\begin{abstract}
A class of discrete nonlinear Schr\"odinger equations with
arbitrarily high order nonlinearities is introduced. These
equations are derived from the same Hamiltonian  using different
Poisson brackets and include as particular cases  the saturable
discrete nonlinear Schr\"odinger equation and the Ablowitz-Ladik
equation. As a common property, these equations possess three
kinds of exact analytical stationary solutions for which the
Peierls-Nabarro barrier is zero. Several properties of these
solutions, including stability, discrete breathers and moving
solutions, are investigated.
\end{abstract}
\maketitle

\section{Introduction}
The discrete nonlinear Schr{\"o}dinger (DNLS) equation appears
ubiquitously \cite{KRB} throughout modern science since it
represents one of the simplest equations in which the combination
of dispersive effects with a cubic nonlinearity leads to localized
solutions of soliton type. Most notable is the role it plays in
understanding the propagation of electromagnetic waves in glass
fibers and other optical waveguides \cite{OP}. More recently the
DNLS  has been used as a tight binding model for Bose-Einstein
condensates in optical lattices \cite{ST,ABDKS}. From a physical
point of view, it is of interest to study the effects of including
high order nonlinear terms (higher than cubic) in the equation on
discrete solitons. These terms appear in different physical
contexts such as Bose gases with hard core interactions in the
Tonks-Girardeau regime \cite{vignolo} and low-dimensional
Bose-Einstein condensates in which quintic nonlinearities in the
nonlinear Schr\"odinger (NLS) equation are used to model
three-body interactions \cite{AS05}. A self-focusing cubic-quintic
NLS equation is also used in nonlinear optics as a model for
photonic crystals \cite{photonic}.

For the continuous NLS equation with attractive interaction it is
well known that the higher order nonlinearities (higher than
cubic) lead to the collapse in a finite time (blow up) if the norm
exceeds a critical value,  even in the one-dimensional case. The
interplay between dimensionality and the order of nonlinearity has
indeed been used in the past as a way to investigate collapse in
low dimensional nonlinear systems \cite{berge}. Although in a
discrete NLS system true collapse cannot occur, due to the
conservation of the norm, it may be possible that some of the
features observed in the continuous NLS system about localized
solutions may also exist at the discrete level. In particular, it
is known that in the 1D continuous NLS equation with high order
nonlinearity (for example quintic) there exists only one localized
solution for each value of the norm (critical norm), the so called
Townes soliton \cite{chiao}, which separates collapsing and decaying 
solutions while being marginally stable against decay or collapse.
In the presence of an external field, for example a periodic
potential, it is possible to stabilize such solutions of
continuous NLS with higher order nonlinearities against decay,
extending the existence range of localized solutions from a single
value of the norm to a whole interval. Since the discrete NLS can
be viewed as a tight binding model of the continuous NLS with a
periodic potential, it is of interest to investigate the existence
of discrete, stable localized solutions when higher order
nonlinearity are introduced in the DNLS.

The aim of this paper is twofold. Firstly, we introduce a general
DNLS equation with arbitrary higher order nonlinearities which in
the appropriate limits reduces to the integrable Ablowitz-Ladik
(AL) equation \cite{AL} and to the cubic DNLS with a saturable
nonlinearity \cite{I}. These two equations have the remarkable property that
they possess different analytical stationary solutions, both periodic
and localized (solitonic), which are exact to all orders of
nonlinearity. These solutions exist for specific values of frequency
and nonlinear parameters and have been shown to be stable under small
perturbations. Secondly, we investigate the effects of higher order
nonlinearity on the stability and mobility of localized solutions
of a generalized DNLS. In particular, we compute the Peierls-Nabarro
(PN) barrier and perform direct numerical integration to show the
existence of moving solutions.

The plan of the paper is as follows. In Sec. II we first show that
the AL model \cite{AL} and the saturable DNLS model \cite{I} can
be obtained from the same Hamiltonian. In Sec. III we extend these
ideas to Hamiltonians with arbitrary nonlinearity and obtain
several higher order DNLS models. We then obtain a number of (time)
stationary, spatially periodic as well as localized solutions. In
Sec. IV we study the various properties of these solutions. 
In Sec. V we examine the question of the existence of moving
solutions in DNLS models with higher order nonlinearities.
Finally in Sec. VI we point out the possible implications of our
results and some open problems.

\section{The model}

In a classic paper, Ablowitz and Ladik \cite{AL} showed that one
of the variants of the DNLS equation given by
\be i\dot{\psi}_n
+(1+|\psi_n|^2)[\psi_{n+1}+\psi_{n-1}] -2\psi_n=0,
\label{EQ:AL}
\ee
is integrable. In Ref. \cite{Salerno} a model was proposed which in
one limit goes over to the DNLS model while in the other limit it
goes over to the integrable AL model. Recently, we were able to
obtain exact periodic solutions of the DNLS equation with a saturable
nonlinearity \cite{I}
\begin{equation}
i\dot{\psi}_n + (\psi_{n+1}+\psi_{n-1}-2\psi_n)+
\frac{\nu|\psi_n|^2}{1+\mu|\psi_n|^2}\psi_n=0, \label{EQ:1}
\end{equation}
which is an established model for optical pulse propagation in
various doped fibers \cite{fibers}. In Eq. (\ref{EQ:1}), $\psi_n$
is a complex valued ``wave function" at site $n$, while $\nu$ and
$\mu$ are real parameters.

We point out that the two equations, i.e.  Eqs. (\ref{EQ:AL})
and (\ref{EQ:1}) can both be derived from the same Hamiltonian
$\cal H$ given by:
\bea
{\cal H}= \sum_{n=1}^{N} \bigg[|\psi_n
-\psi_{n+1}|^2-\frac{\nu}{\mu}|\psi_n|^2
\nonumber
\eea
\bea
+\frac{\nu}{\mu^2} \ln \left (1+\mu|\psi_n|^2\right ) \bigg],
\label{EQ:HAM}
\eea
\noindent
and the equation of motion in both cases is
\begin{equation}
i \dot \psi_n =[\psi_n,\cal H]\,.
\label{EQ:EQ_0}
\end{equation}
\noindent The difference in the equations of motion comes
from a different definition of the Poisson bracket (PB) and
consequently a different definition of the time derivative. The
Poisson bracket structure in both the cases can  be compactly
written as
\begin{equation}
[U,V]=\sum_{n=1}^{N}
(\frac{\partial U}{\partial \psi_n}\frac{\partial V}{\partial \psi_n^*}
-\frac{\partial U}{\partial \psi_n^*}\frac{\partial V}{\partial \psi_n})
(1+\lambda|\psi_n|^2)\,.
\label{EQ:POS_L}
\end{equation}
\noindent On using Eqs. (\ref{EQ:HAM}) and Eq. (\ref{EQ:EQ_0}) for
$\lambda=0$ ($\psi_n$ and $i\psi_n^*$ are
conjugate variables) it yields Eq. (\ref{EQ:1}) \cite{I} through Eq.
(\ref{EQ:EQ_0}), while $\lambda=\mu$ ($\psi_n$ and $i\psi_n^*$ are
non-conjugate variables) yields the equation introduced in Ref. \cite{Salerno}
\begin{equation}
i\dot{\psi}_n + (1+\mu|\psi_n|^2)(\psi_{n+1}+\psi_{n-1}-2\psi_n)+
\nu|\psi_n|^2\psi_n=0. \label{EQ:SAL}
\end{equation}
Notice that for $\mu\rightarrow 0$ both Eq. (\ref{EQ:1}) and Eq.
(\ref{EQ:SAL}) reduce to the ordinary DNLS. Therefore, in the
following we will assume $\mu\neq 0$ and perform the transformation
$\sqrt{\mu}\psi_n\rightarrow\psi_n$. This will replace $\mu$ by 1 and
$\frac{\nu}{\mu}$ by $\nu$, thus rendering the problem a one parameter
problem.\\

For the $\lambda=0$ case \cite{I} the equation of motion, Eq.
(\ref{EQ:1}), can be written as
\bea i(1+|\psi_n|^2)\dot{\psi}_n +
(1+|\psi_n|^2)(\psi_{n+1}+\psi_{n-1}-2\psi_n)
\nonumber
\eea
\bea
+ \nu|\psi_n|^2\psi_n=0,
\label{EQ:EQ_1}
\eea
\noindent while for the $\lambda=1(=\mu)$ Eq. (\ref{EQ:SAL}) becomes:
\begin{equation}
i\dot{\psi}_n +
(1+|\psi_n|^2)(\psi_{n+1}+\psi_{n-1}-2\psi_n)+
\nu|\psi_n|^2\psi_n=0.
\label{EQ:SAL2}
\end{equation}
\noindent Note that $\nu=2$ in Eq. (\ref{EQ:SAL2}) gives the AL
equation \cite{AL}, Eq. (\ref{EQ:AL}).\\

In both cases a conserved {\em power} ${\cal P}$ can be written:\\
\noindent
\begin{equation}
{\cal P}=\sum_{n=1}^{N}\frac{1}{\lambda} \ln(1+\lambda|\psi_n|^2)\,,
~~\lambda \rightarrow 0~~\mbox{or} ~~\lambda = \mu =1\,.
\label{eq:POWER_M}
\end{equation}
\noindent
The difference between the two cases is the presence of $i|\psi_n|^2$
in the factor in front of the time derivative term in Eq. (\ref{EQ:EQ_1}).
However, for a stationary solution [i.e. only $\exp(-i\omega t)$ time
dependence: $i\dot{\psi}_n=\omega\psi_n$] Eq. (\ref{EQ:EQ_1}) can be
written as
\bea
i\dot{\psi}_n +
(1+|\psi_n|^2)(\psi_{n+1}+\psi_{n-1}-2\psi_n)
\nonumber
\eea
\bea
+ (\nu+\omega)|\psi_n|^2\psi_n=0\,,
\label{EQ:EQ_1.1}
\eea
\noindent
which is identical in form to Eq. (\ref{EQ:SAL}).\\
\noindent

The exact solutions to Eq. (\ref{EQ:EQ_1}) given in \cite{I} all are
stationary solutions ``rotating" with the frequency
\begin{equation}
\omega=2-\nu\,.
\label{EQ:Frequency}
\end{equation}
\noindent Inserting this frequency into Eq. (\ref{EQ:EQ_1.1})
gives the AL equation:
\begin{equation}
i\dot{\psi}_n +
(1+|\psi_n|^2)(\psi_{n+1}+\psi_{n-1}-2\psi_n)+
2|\psi_n|^2\psi_n=0\,.
\label{EQ:AL2}
\end{equation}
\noindent  From this is clear that exact stationary solutions
of the saturable DNLS equation \cite{I} are also stationary
solutions of the AL equation. Analytical expressions for
stationary and moving solutions of the AL equation were given by
Scharf and Bishop in Ref. \cite{Alan} (note that their $\omega$
differs from ours by a factor of two, see Eq.
(\ref{EQ:Frequency}). We also remark that, because of the
frequency relation $\omega+\nu =2$, Eq. (\ref{EQ:SAL}) has
stationary solutions of the AL only if $\nu=2$, i.e. when it
reduces to the AL equation. In the following we shall demonstrate
the existence of analytical stationary solutions of the AL type
also for generalizations of Eqs. (\ref{EQ:1}), (\ref{EQ:SAL}),
with arbitrarily high order nonlinearities.

\subsection{High order nonlinearities}

\noindent We now generalize the nonlinear part of the Hamiltonian
$\cal H$ in Eq. (\ref{EQ:HAM}) by replacing $\ln(1+|\psi_n|^2)$ by:
\be \ln(1+f(|\psi_n|^2)), \ee
where the function $f(x)$, a polynomial of degree $p+1$, is given by:
\be f(x)=\sum_{j=0}^{p}
\alpha_jx^{j+1}.
\ee
Here $\alpha_0$ is always 1. Having already considered the case
$\mu\rightarrow 0$, and having chosen $\mu=1$, we can next choose
two different values of $\nu$ in Eq. (\ref{EQ:HAM}). Therefore,
instead of Eq. (\ref{EQ:HAM}), the Hamiltonian ${\cal H}$ is now
given by:
\bea {\cal H} = \sum_{n=1}^{N} \bigg[|\psi_n
- \psi_{n+1}|^2 -\nu'|\psi_n|^2
\nonumber
\label{polyn}
\eea
\bea
+\nu\ln(1+
f(|\psi_n|^2)) \bigg],
\label{Ham:gen}
\eea
and the generalized Poisson bracket by:
\bea
[U,V]=
\nonumber
\eea
\bea
\sum_{n=1}^{N} \left (\frac{dU}{d\psi_n}\frac{dV}{d \psi_n^*}-
\frac{dV}{d\psi_n}\frac{dU}{d \psi_n^*}\right ) \left [1+ \lambda
f(|\psi_n|^2) \right ],
\label{PB}
\eea
with $\lambda = 0$ or 1.  Note that the equation of motion is still
given by Eq.  (\ref{EQ:EQ_0}).

\noindent
We notice for the $\lambda = 0$ case that a transformation $\psi_n
\rightarrow e^{-i\Delta\omega t}\psi_n$ would add $\Delta\omega$ to
the coefficient of $\psi_n$ in the equation of motion and subtract
the same from the coefficient of the $|\psi_n|^2$ term in the
Hamiltonian.  Therefore, a $\nu'$ different from $\nu$ will only
shift the frequency. We remark that some of the recently discussed
models \cite{had} fall in this category and are essentially equivalent
to the saturated DNLS model \cite{I}. Thus, both in this section and
in Sec. IIIA the effect will only be to change the frequency $\omega$
by $\Delta\omega$.  This will, however, not be the case for $\lambda
= 1$ as discussed here and in Sec. IIIB.\\

\noindent
We also note that with higher order nonlinearities too, besides
the Hamiltonian ${\cal H}$, the {\em power}
\be\label{p1}
{\cal P} = \sum_{n=1}^{N}|\psi_n|^2\,,~~\lambda=0\,,
\ee
or
\be\label{p2}
{\cal P} = \sum_{n=1}^{N}\ln(1+ f(|\psi_n|^2))\,,~~\lambda=1\,,
\ee
is a conserved quantity.\\

\section{Exact analytical solutions}
\noindent
The main objective here is to find stationary solutions of the type
obtained in Ref. \cite{I} but with an additional amplitude factor
$A$ in the equation of motion derived from the generalized Hamiltonian
${\cal H}$ given by Eq. (\ref{Ham:gen}).  In particular, as in Ref.
\cite{I} we try to obtain three different types of solutions. The type
I solution is given by
\be
\psi_n^I =
A\frac{\mbox{sn}(\beta,m)}{\mbox{cn}(\beta,m)}
e^{-i(\omega t+\delta)} \dn(\beta(n+c),m)\,,
\label{SOL:DN,o}
\ee
where the frequency $\omega$ is given by Eq. (\ref{EQ:Frequency}) while
the two equations determining $m$ and $\beta$ are
\be
\beta=\frac{2K(m)}{N_p}\,,~~
\nu=2\frac{\dn(\beta,m)}{\cn^2(\beta,m)}\,.
\label{SOL:DN,o,con}
\ee
Here $N_p$ denotes the number of sites in one period of the system.
On the other hand, the type II solution is given by
\be
\psi_n^{II} =
A\sqrt{m}\frac{\mbox{sn}(\beta,m)}{\mbox{dn}(\beta,m)}
e^{-i(\omega t+\delta)} \cn(\beta(n+c),m),
\label{SOL:CN,o}
\ee
with the same frequency $\omega$ as given by Eq. (\ref{EQ:Frequency})
and the two equations  determining $m$ and $\beta$ being:
\be
\beta=\frac{4K(m)}{N_p}\,,~~
\nu=2\frac{\cn(\beta,m)}{\dn^2(\beta,m)}\,,
\label{SOL:CN,o,con}
\ee
Here, $\sn(x,m)$, $\cn(x,m)$, and $\dn(x,m)$ are the Jacobi elliptic
functions of modulus $m$, while $K(m)$ is the complete elliptic
integral of the first kind \cite{stegun}.
The two solutions have a common limit for $m \rightarrow 1$ giving
the type III solution:
\be
\psi_n^{III} =
A\frac{\sinh(\beta)}{\cosh(\beta (n+c))}e^{-i(\omega t+\delta)},
~(N_p\rightarrow\infty)\,,
\label{SOL:SECH,o}
\ee
with $\beta$ being determined by:
\be
\nu=2\cosh (\beta)\,.
\label{SOL:SECH,o,con}
\ee
\noindent

\subsection{Standard Poisson brackets, $\lambda =0$}

\noindent
In this case of the standard PB ($\lambda =0$), the equation of motion
becomes:
\bea
i\dot \psi_n +\left [\psi_{n+1}-2\psi_n+\psi_{n-1}\right ]
\nonumber
\eea
\bea
+\frac{\nu'(1+f(|\psi_n|^2)-\nu f'(|\psi_n|^2))}{(1+f(|\psi_n|^2))}\psi_n
=0\,,
\eea
or equivalently,
\bea
(i\dot \psi_n -2\psi_n+\nu'\psi_n)(1+f(|\psi_n|^2)
\nonumber
\eea
\bea
+\left [\psi_{n+1}+\psi_{n-1}\right ](1+f(|\psi_n|^2))
-\nu f'(|\psi_n|^2)\psi_n
=0\,.
\eea
This equation has the stationary solutions of the form (Type I, II, and
III) as given by Eqs. (\ref{SOL:DN,o}), (\ref{SOL:CN,o}), and
(\ref{SOL:SECH,o}) where the frequency $\omega$ is still given by Eq.
(\ref{EQ:Frequency}) but with $\nu$ replaced by $\nu'$:
\begin{equation}
\omega=2-\nu'\,.
\label{EQ:Frequency2}
\end{equation}
Also, the two equations needed to determine $m$ and $\beta$ are still
the same. The condition on the amplitude is that $A=\sqrt{p+1}$ and
that the coefficients of the polynomial $\alpha_j$ in Eq. (\ref{polyn})
are given by
\be
\alpha_j = \frac{p}{(j+1)!}
\frac{\Pi_{k=1}^{j-1} (p-k)}{(p+1)^{j}},~~1\le j\le p\,.
\label{COEFF}
\ee
This specifies the equation completely.  We find:
\be
1+f(x)=\left(1+\frac{x}{p+1}\right)^{p+1} ~~\rightarrow e^x
~\mbox{for}~ p\rightarrow\infty ~~.
\ee
\noindent
For the first few values of $p$ one finds:
\[
\begin{array}{lllll}
p=1:& \alpha_1=\frac{1}{4},&~&~&~\\
p=2:& \alpha_1=\frac{1}{3},&\alpha_2=\frac{1}{27},&~&~\\
p=3:& \alpha_1=\frac{3}{8},&\alpha_2=\frac{1}{16},
&\alpha_3=\frac{1}{256},&~\\
p=4:& \alpha_1=\frac{2}{5},&\alpha_2=\frac{2}{25},
&\alpha_3=\frac{1}{125},&\alpha_4=\frac{1}{3125}.\\
\end{array}
\]
\noindent
We note that $p=0$ gives the results of Ref. \cite{I}. We also note that
the $p>0$ cases are just a rescaling of the results of Ref. \cite{I}.
Thus the higher order nonlinearities do not give anything essentially
new in the $\lambda=0$ case.

\subsection{Non-standard Poisson brackets, $\lambda= 1$}

\noindent This case of the non-standard PB represents a
generalization of Eq. (\ref{EQ:SAL}). From Eq. (\ref{EQ:EQ_0}) and
Eq. (\ref{PB}) we get
\bea
i \dot \psi_n+\left  [\psi_{n+1}+\psi_{n-1} \right ]
[1+f(|\psi_n|^2)]
\nonumber
\eea
\bea
+(\nu' - 2)[1+f(|\psi_n|^2)]\psi_n
-\nu f'(|\psi_n|^2)\psi_n
 =0\,.
\eea In order to use the identities for the Jacobi elliptic
functions we must choose $\nu' =2$, and we get a restricted
generalization of Eq. (\ref{EQ:SAL})

\bea i \dot \psi_n+\left [\psi_{n+1}+\psi_{n-1} \right ]
[1+f(|\psi_n|^2)] \nonumber \eea \bea -\nu f'(|\psi_n|^2)\psi_n
 =0\,.
\label{Gen_Salerno_1}
\eea
This equation has the stationary solutions of the form (Type I, II, and
III) as given by Eqs. (\ref{SOL:DN,o}), (\ref{SOL:CN,o}), and
(\ref{SOL:SECH,o}) except here the equations determining $m$ and
$\beta$, are
\be
\beta=\frac{2K(m)}{N_p}\,,~~
\nu-\omega=2\frac{\dn(\beta,m)}{\cn^2(\beta,m)}\,
\label{SOL:DN,o,con,l=1}
\ee
for the type I solutions,
\be
\beta=\frac{4K(m)}{N_p}\,,~~
\nu-\omega=2\frac{\cn(\beta,m)}{\dn^2(\beta,m)}\,
\label{SOL:CN,o,con,l=1}
\ee
for type II, and finally 
\be
\nu-\omega=2\cosh (\beta)\,
\label{SOL:SECH,o,con,l=1}
\ee
for type III.
The
condition on the amplitude is now that $A=\sqrt{p+1}\sqrt{\frac{\nu}{\nu
-\omega}}$ and that the coefficients of the polynomial $\alpha_j$ are
given by
\bea
\alpha_j = \frac{1}{(j+1)!}
\frac{\Pi_{k=1}^{j-1} (p-k)}{(p+1)^{j-1}}
\left(\frac{\nu-\omega}{\nu}\right)^j\left(1-\frac{\nu-\omega}{\nu(p+1)}
\right)\,,
\nonumber
\eea
\bea
1\le j\le p\,.
\label{COEFF2}
\eea
Note that the frequency $\omega$ now also appears as a parameter.
This completely specifies our equation.

\noindent
We find:
\bea
1+f(x)=\frac{1}{p}\left(\frac{\nu(p+1)}{\nu-\omega}-1\right)
\left(1+\frac{\nu-\omega}{\nu}\frac{x}{p+1}\right)^{p+1}
\nonumber
\eea
\bea
-\frac{\omega}{\nu-\omega}\frac{p+1}{p}\left(1+\frac{\nu-\omega}{\nu}
\frac{x}{p+1}\right)\,.
\eea
In addition, for $p\rightarrow\infty$
\be
1+f(x)\rightarrow\frac{\nu}{\nu-\omega}e^{\frac{\nu-\omega}{\nu}x}
-\frac{\omega}{\nu-\omega}\,,
\label{limit1}
\ee
and
\be
f'(x)\rightarrow e^{\frac{\nu-\omega}{\nu}x}\,.
\label{limit2}
\ee
\noindent
For the first few values of $p$ we find:\\
\[
\begin{array}{llll}
p=1:& \alpha_1 =\frac{\nu-\omega}{2\nu}(1-\frac{\nu-\omega}{2\nu}),&~&\\
p=2:& \alpha_1 =\frac{\nu-\omega}{2\nu}(1-\frac{\nu-\omega}{3\nu}),&
\alpha_2 =\frac{(\nu-\omega)2}{18\nu2}(1-\frac{\nu-\omega}{3\nu}),&\\
p=3:& \alpha_1 =\frac{\nu-\omega}{2\nu}(1-\frac{\nu-\omega}{4\nu}),&
\alpha_2 =\frac{(\nu-\omega)2}{12\nu2}(1-\frac{\nu-\omega}{4\nu}),&\\
~&\alpha_3 =\frac{(\nu-\omega)3}{192\nu3}(1-\frac{\nu-\omega}{4\nu}).\\
\end{array}
\]
\noindent
Note that if $\omega=0$ (i.e., $\frac{\nu-\omega}{\nu}=1$) then, as
expected, $\alpha_j$ are the same in non-standard Poisson bracket case
as in the standard Poisson bracket case.  We would like to remind that
if one also chooses $\nu =2$ then one obtains the AL model and its
higher order generalizations.

\noindent
It may be noted here that in order to obtain the above solutions we
have made use of  two identities for the Jacobi elliptic functions
\cite{Khare}.  The first identity is
\bea
\label{dni}
&&\dn^2 (x)[\dn(x+a)+\dn(x-a)]= A \dn (x)
\nonumber \\
&&+B[\dn(x+a)+\dn(x-a)]\,,
\eea
where
\be
\label{dnj}
A=2\ns(a)\ds(a)\,,~~B=-\cs^2 (a)\,.
\ee
Here $\ns(a) \equiv \frac{1}{\sn(a)}, \ds(a) \equiv \frac{\dn(a)}{\sn(a)},
\cs(a) \equiv \frac{\cn(a)}{\sn(a)}$. We have suppressed the modulus $m$
in our notation here.  On repeatedly multiplying both sides of the
identity (\ref{dni}) by $\dn^2(x)$ and simplifying yields the following
general identity of arbitrary (odd) rank:
\bea\label{dna}
&&\dn^{2n} (x)[\dn(x+a)+\dn(x-a)]=
B^{n} [\dn(x+a) \nonumber \\
&&+\dn(x-a)]+A \sum_{j=1}^{n} B^{j-1}
\dn^{2(n-j)+1} (x)\,,
\eea
which has been used in deriving some of the above solutions. Here $A$,
$B$ are the same as given by Eq. (\ref{dnj}).

The second identity we have used above is \cite{Khare}
\bea\label{cni}
&&m \cn^2 (x)[\cn(x+a)+\cn(x-a)]= A \cn (x) \nonumber \\
&&+B[\cn(x+a)+\cn(x-a)]\,,
\eea
where
\be\label{cnj}
A=2\ns(a)\cs(a)\,,~~B=-\ds^2 (a)\,.
\ee
On repeatedly multiplying both sides of this identity by $m\cn^2(x)$
and simplifying yields the following general identity of arbitrary (odd)
rank:
\bea\label{cna}
&&m^{n}\cn^{2n} (x)[\cn(x+a)+\cn(x-a)]=
B^{n} [\cn(x+a) \nonumber \\
&&+\cn(x-a)] +A \sum_{j=1}^{n} m^{n-j} B^{j-1}
\cn^{2(n-j)+1} (x)\,,
\eea
which has been used in deriving some of the above solutions. Here $A$,
$B$ are the same as given by Eq. (\ref{cnj}).

\subsection{The $p\rightarrow\infty$ limit and the linear limit}

\noindent
Both $\lambda=0$ and $\lambda=1$ cases have the same linear (small signal) limit:
\be\label{lin}
i\dot\psi_n-(2+\nu-\nu')\psi_n +(\psi_{n+1}+\psi_{n-1})=0.
\ee
For the $\lambda=0$ case, the $p\rightarrow\infty$ limit gives the same
equation (\ref{lin}), because in this limit $1+f(x)=f'(x)=e^x$.
For the $\lambda=1$
case, however,  $1+f(x)$ and $f'(x)$ differ as is clear from
Eqs. (\ref{limit1}) and (\ref{limit2}).
Inserting these two equations into Eq. (\ref{Gen_Salerno_1}) yields
\bea
i\dot\psi_n-\omega\frac{\psi_{n+1}+\psi_{n-1}}{\nu-\omega}
\nonumber
\eea
\bea
+ \nu e^{\frac{\nu-\omega}{\nu}x}\left(\frac{\psi_{n+1}+\psi_{n-1}}
{\nu-\omega}-\psi_n\right) =0.
\eea
This equation is a little bit tricky since it contains terms of the
order $\sqrt{p}$ and $\sqrt{p}e^p$. We must balance the $\sqrt{p}e^p$
terms first giving:
\be
\psi_{n+1}+\psi_{n-1}=(\nu-\omega)\psi_n,
\ee
and next balance the smaller terms giving
\be
i\dot\psi_n=\omega\psi_n.
\ee
Note that in this case we end up with two coupled linear equations,
one describing the spatial variation and one describing the temporal
variation.

\section{Properties of the New Solutions}

\subsection{PN Barrier}

In a discrete lattice there is an energy cost associated with moving
localized modes (such as a soliton or a breather) by a half lattice
constant. This is the celebrated Peierls-Nabarro (PN) barrier
\cite{PN,peyrard}.  As is well known, while for the AL
(i.e. $p=0,\lambda=1,\nu=2$) case, this barrier is known to be zero,
as shown by us in Ref. \cite{I}, this barrier is nonzero in the case
of the saturated DNLS model (i.e.  $p=0$,
$\lambda=0$). It is then of significant interest to know whether this
barrier exists in models with higher order nonlinearities.  Since for
$p=0$, $\lambda=0$, we have already studied the various properties of
the solutions in Ref. \cite{I} and since higher order nonlinearities
do not give anything essentially new, we will only study the properties
of the $\lambda=1$ solutions. In particular, we will show that like the
AL case, even for all the higher order models, the PN barrier is zero.

In view of $\nu'=2$, the Hamiltonian (\ref{Ham:gen}) (for $\lambda=1$)
takes a simple form
\bea
{\cal H} = \sum_{n=1}^{N} \bigg[-\psi_n \psi_{n+1}^{*}-
 \psi_n^{*} \psi_{n+1}
+\nu\ln(1+ f(|\psi_n|^2)) \bigg]
\nonumber
\eea
\bea
\equiv H_1+H_2\,,
\label{pn1}
\eea
while the power $P$ is as given by Eq. (\ref{p2}). The first part
of the Hamiltonian without the logarithmic term ($H_1$) is easily
evaluated in the case of all the three solutions as given by Eqs.
(\ref{SOL:DN,o}), (\ref{SOL:CN,o}) and (\ref{SOL:SECH,o}) and it
is easily shown that for all the three solutions the answer is
independent of the constant $c$, i.e., the distance from a lattice
point where the center of the elliptic soliton solution is located.
For example, for solution of type I, we obtain
\be
H_1^{I} =-\frac{2A^2 N_{p}}{\cs^2 (\beta,m)} [\dn (\beta,m)-
\cs(\beta,m) Z(\beta,m)]\,,
\ee
while for the solution of type II
\be
H_1^{II} =-\frac{2A^2 N_{p}}{\cs^2 (\beta,m)} [m\cn (\beta,m)-
\ds(\beta,m) Z(\beta,m)]\,,
\ee
and for solution of type III
\be
H_1^{III} =-4A^2 \sinh (\beta)\,.
\ee
Here $Z(\beta,m)$ is the Jacobi zeta function \cite{stegun}.
While deriving these relations, use has been made of the identities
in Ref. \cite{Khare}
\be
\dn(x)\dn(x+a)=\dn(a)-\cs(a)Z(a)+\cs(a)[Z(x+a)-Z(x)]\,,
\ee
\bea
&&m\cn(x)\cn(x+a)=m\cn(a)-\ds(a)Z(a) \nonumber \\
&&+\ds(a)[Z(x+a)-Z(x)]\,,
\eea

We are unable to evaluate the expression for power $P$ and (hence) the
second term in the Hamiltonian $H_2$ analytically for any nonzero $p$
in the case of solutions of type I and II. However, this is easily
accomplished in the case of the localized solutions (of type III).
However, even without the explicit computation, it is easy to show
that the PN barrier is zero in the case of all three solutions.  Let
us first explain the key idea.  The power $P$ is given by
\be
P=\sum_{n=1}^{N} \ln [1+\phi_n^2+\alpha_1\phi_n^4+...+\alpha_p
\phi_n^{2p+2}]\,,
\ee
where $\psi_n=\phi_n e^{-i(\omega t+\delta)}$ and $\phi_n$ is easily
obtained from Eqs. (\ref{SOL:DN,o}), (\ref{SOL:CN,o}) and
(\ref{SOL:SECH,o}).  The key point to note is that the expression under
the logarithm can always be factorized as
\be
\left(1+\frac{\nu-\omega}{(p+1)\nu}\phi_n^2\right)\left[1+\frac{2\alpha_1
\nu}{\nu-\omega}
\phi_n^2+...+\frac{(p+1)\alpha_p \nu}{\nu-\omega}\phi_n^{2p}\right]\,,
\ee
which can be further factorized as
\be
\left(1+\frac{\nu-\omega}{(p+1)\nu}\phi_n^2\right)[1+a_1\phi_n^2][1+a_2
\phi_n^2]...[1+a_p\phi_n^{2}]\,,
\ee
where $a_1,a_2,...,a_p$ are the roots of the above equation. For example
\be
\sum_{j=1}^{p} a_j = \frac{2\alpha_1 \nu}{\nu-\omega}\,,~~
\prod_{j=1}^{p} a_j =\frac{(p+1)\alpha_p \nu}{\nu-\omega}\,.
\ee
Hence the expression for power takes the simple form
\bea
&&P=\sum_{n=1}^{N} \ln \left[1+\frac{\nu-\omega}{(p+1)\nu}\phi^2\right]
\nonumber \\
&&+\sum_{j=1}^{p} \sum_{n=1}^{N} \ln [1+a_j \phi_n^2]\,.
\eea

We now observe that
in the celebrated AL case, the power $P$ is
given by
\be
P=\sum_{n=1}^{N} \ln [1+\phi_n^2]\,,
\ee
where $\phi_n$ is either $a \dn [\beta(n+c-vt,m)]$ or $\dn$ replaced by
$\cn [\beta(n+c-vt,m)]$ or by $\sech [\beta(n+c-vt)]$, and it is well
known in that case that this sum is independent of $c$ since $P$ is a
constant of motion and $t$ and $c$ always come together in this
expression \cite{cai}. As a result, it immediately follows that even
for the higher order DNLS models the power $P$ and hence $H_2$ must
also be $c$-independent being a sum of $p$ terms of the form same as
that appears in the AL case.  Thus we see that remarkably enough, even
in higher order DNLS models the PN barrier is zero in the case of all
three solutions.

Unfortunately for the solutions of type I and II we are unable to
perform the additions analytically and hence compute $P$ and $H_2$
analytically.  However, for the spatially localized solutions of type
III this is easily accomplished.  We observe that for the localized
solutions of type III, each term in the sum has the form
\be
\sum_{n=-\infty}^{\infty } \ln [1+a\sech^2 \beta(n+c)]\,,
\ee
and as is well known from the AL case \cite{cai}, this sum is
$c$-independent and given by
\be
\sum_{n=-\infty }^{\infty } \ln \left[1+a\sech^2 \beta(n+c)\right]
=\frac{2}{\beta} \left[\sinh^{-1} \sqrt{a}\right]^2 \,.
\ee
Thus, in principle we know $P$ and $H_2$ for type III solutions.
As an illustration, for the $p=1$ case, it is easily shown that
\bea
&&H_2=\nu P =2\beta \nu \nonumber \\
&&+\frac{2\nu}{\beta} \bigg [\sinh^{-1} \left(\sqrt{\frac{2\nu}{\nu
-\omega}[1-\frac{\nu-\omega}{2\nu}]}\sinh \beta\right) \bigg ]^2\,.
\eea
Generalizing, for arbitrary $p$, the power P and $H_{2}$ are given by
\bea
&&H_2=\nu P +2\beta \nu \nonumber \\
&&+\frac{2\nu}{\beta}\sum_{j=1}^{p} \left[\sinh^{-1}\left(\sqrt{(p+1)
\frac{\nu}{\nu-\omega}a_j}\sinh \beta\right)\right]^{2}\,.
\eea
It is indeed remarkable that the PN barrier is not only zero for the AL
model but for even higher order generalizations. It would be worthwhile
to examine if the higher order models are also integrable, although
perhaps the answer is likely not in the affirmative.

\subsection{Stability}
\noindent
In order to study the linear stability of the exact solutions $\psi_n^j$
($j$ is I, II, or III) we introduce the following expansion
\begin{equation}
\psi_n(t)=\psi_n^j+\delta \psi_n(t) e^{-i\omega t},
\label{EQ:STAB_1}
\end{equation}
applied in a frame rotating with frequency $\omega$ of the solution.
The stability analysis for the $p=0,\lambda = 0$ case was carried out
in Ref. \cite {I} and as seen above, higher order nonlinearities do not
give any new solution. We will therefore only consider the stability of
$\lambda=1$ solutions.  Upon using this expansion into the equation of
motion (for $\lambda = 1, \nu' = 2$)
\bea
i\dot{\psi}_n
+[\psi_{n+1}+\psi_{n-1}][1+f(|\psi_n|^2)]
\nonumber
\eea
\bea
-\nu f'(|\psi_n|^2)\psi_n =0.
\eea
Next, retaining only terms linear in the perturbation, and taking into
account the basic frequency $\omega$ of the unperturbed solutions and
the perturbations, we get
\begin{eqnarray}
&&i\delta\dot{\psi}_n+\big(\delta \psi_{n+1}+\delta
\psi_{n-1}\big)[1+f(|\psi_n|^2)]\nonumber\\
&&+(\omega -\nu f'(|\psi_n|^2))\delta \psi_n
+\big(\psi_n^*(\psi_{n+1}+\psi_{n-1})f'(|\psi_n|^2)\nonumber\\
&&-\nu f''(|\psi_n|^2)|\psi_n|^2\big)
(\delta\psi_n+\delta\psi_n^*)
=0.
\label{EQ:STAB_LIN}
\end{eqnarray}
Continuing by splitting the perturbation $\delta \psi_n $ into real parts
$\delta u_n$
and imaginary parts $\delta v_n$ ($\delta \psi_n =\delta u_n+i\delta v_n$)
and introducing the two real vectors
\begin{eqnarray}
\delta\mbox{\boldmath $U$}=\{\delta u_n\}&~~\mbox{and}~~&
\delta\mbox{\boldmath $V$}=\{\delta v_n\}
\end{eqnarray}
and the two real matrices $\mbox{\boldmath $A$}=\{A_{nm}\}$ and
$\mbox{\boldmath $B$}=\{B_{nm}\}$ by defining
\begin{eqnarray}
&&A_{nm}=
[\delta_{n,m+1}+\delta_{n,m-1}][1+f(|\psi_n|^2)]
\nonumber \\
&&+\bigg (\omega-\nu f'(|\psi_n|^2)-2\nu f''(|\psi_n|^2)|\psi_n|^2
\nonumber \\
&&+2\psi_n^*[\psi_{n+1} +\psi_{n-1}]f'(|\psi_n|^2)
\bigg ) \delta_{nm}\,,
\nonumber \\
&&B_{nm}=
[\delta_{n,m+1}+\delta_{n,m-1}][1+f(|\psi_n|^2)]
\nonumber \\
&& +[\omega-\nu f'(|\psi_n|^2)]\delta_{nm}\,,
\end{eqnarray}
where $m\pm 1$ in the Kronecker $\delta$ means: $m\pm 1~mod~N$.  Then Eq.
(\ref{EQ:STAB_LIN}) can be written compactly as
\begin{eqnarray}
-\delta\mbox{\boldmath $\dot{V}$}+
\mbox{\boldmath $A$}\delta\mbox{\boldmath $U$}=\mbox{\boldmath $0$},&\mbox{and}&
\delta\mbox{\boldmath $\dot{U}$}+
\mbox{\boldmath $B$}\delta\mbox{\boldmath $V$}=\mbox{\boldmath $0$},
\end{eqnarray}
where an overdot denotes time derivative.
Combining these first order differential equations we get:
\begin{eqnarray}
\delta\mbox{\boldmath $\ddot{V}$}+\mbox{\boldmath $A$}\mbox{\boldmath $B$}
\delta\mbox{\boldmath $V$}=\mbox{\boldmath $0$},&\mbox{and}&
\delta\mbox{\boldmath $\ddot{U}$}+\mbox{\boldmath $B$}\mbox{\boldmath $A$}
\delta\mbox{\boldmath $U$}=\mbox{\boldmath $0$}.
\end{eqnarray}
The two matrices $\mbox{\boldmath $A$}$ and $\mbox{\boldmath $B$}$ are
symmetric and have real elements. However, since they do not commute
$\mbox{\boldmath $A$}\mbox{\boldmath $B$}$ and
$\mbox{\boldmath $B$}\mbox{\boldmath $A$}=
(\mbox{\boldmath $A$}\mbox{\boldmath $B$})^{T}$
($T$ means transpose) are not symmetric.
$\mbox{\boldmath $A$}\mbox{\boldmath $B$}$ and
$\mbox{\boldmath $B$}\mbox{\boldmath $A$}$ have the same eigenvalues,
but different eigenvectors. The eigenvectors for each of the two
matrices need not be orthogonal.  The eigenvalue spectrum $\{\gamma \}$
of the matrices $\mbox{\boldmath $A$}\mbox{\boldmath $B$}$ and
$\mbox{\boldmath $B$}\mbox{\boldmath $A$}$ determines the stability of
the exact solutions. If it contains negative eigenvalues, the solution
is unstable. The eigenvalue spectrum always contains two eigenvalues
which are zero. These eigenvalues correspond to the translational
invariance ($c$) and to the invariance of the solution $\psi_n^j$ to
a constant phase factor $e^{-i\delta}$ (i.e., translation in time),
respectively.

In Fig.1 we show three examples of such stability evaluation. Figure 1
shows the lowest (non-zero) eigenvalue from the spectra of type III
solutions obtained for $p=0$ (solid line) , $p=1$ (dashed line), and $p=2$
(dashed-dotted line).  For the integrable AL case ($p=0$) we observe the expected
result that all eigenvalues have a positive real value, indication that
the exact solution is stable for all widths $\beta$. In contrast, we see
that for both $p=1$ and $p=2$ the solution becomes unstable for certain
values of $\beta$. This instability occurs for relatively large values
of $\beta$, and thus when the solutions are very localized.

\begin{figure}\centerline{
\includegraphics[width=\columnwidth,angle=0,clip]{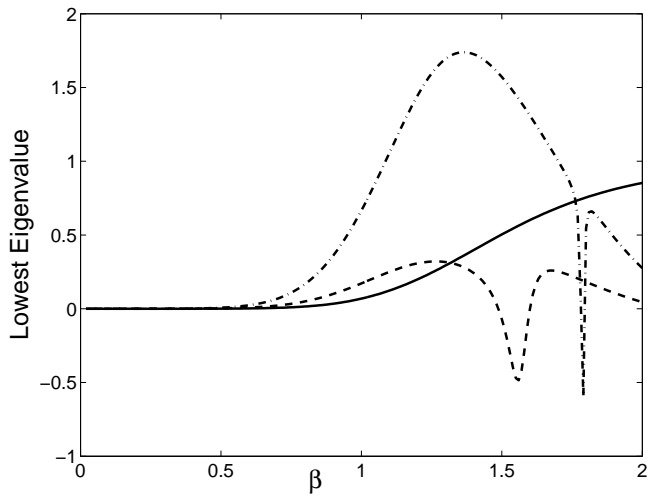}}
\caption{Lowest non zero-eigenvalues for type III solutions [Eq. (23)]
for the (solid line) integrable AL equation (p=0), and for the
nonintegrable cases of p=1 (dashed line), and p=2 (dashed-dotted line). For 
$p=1$ and $p=2$, the eigenvalues are observed to become negative,
indicating instability, for certain values of $\beta$. }
\label{fig0}
\end{figure}

\section{Discrete breathers and moving solitons}
\begin{figure}\centerline{
\includegraphics[width=4.2cm,height=4.2cm,angle=0,clip]{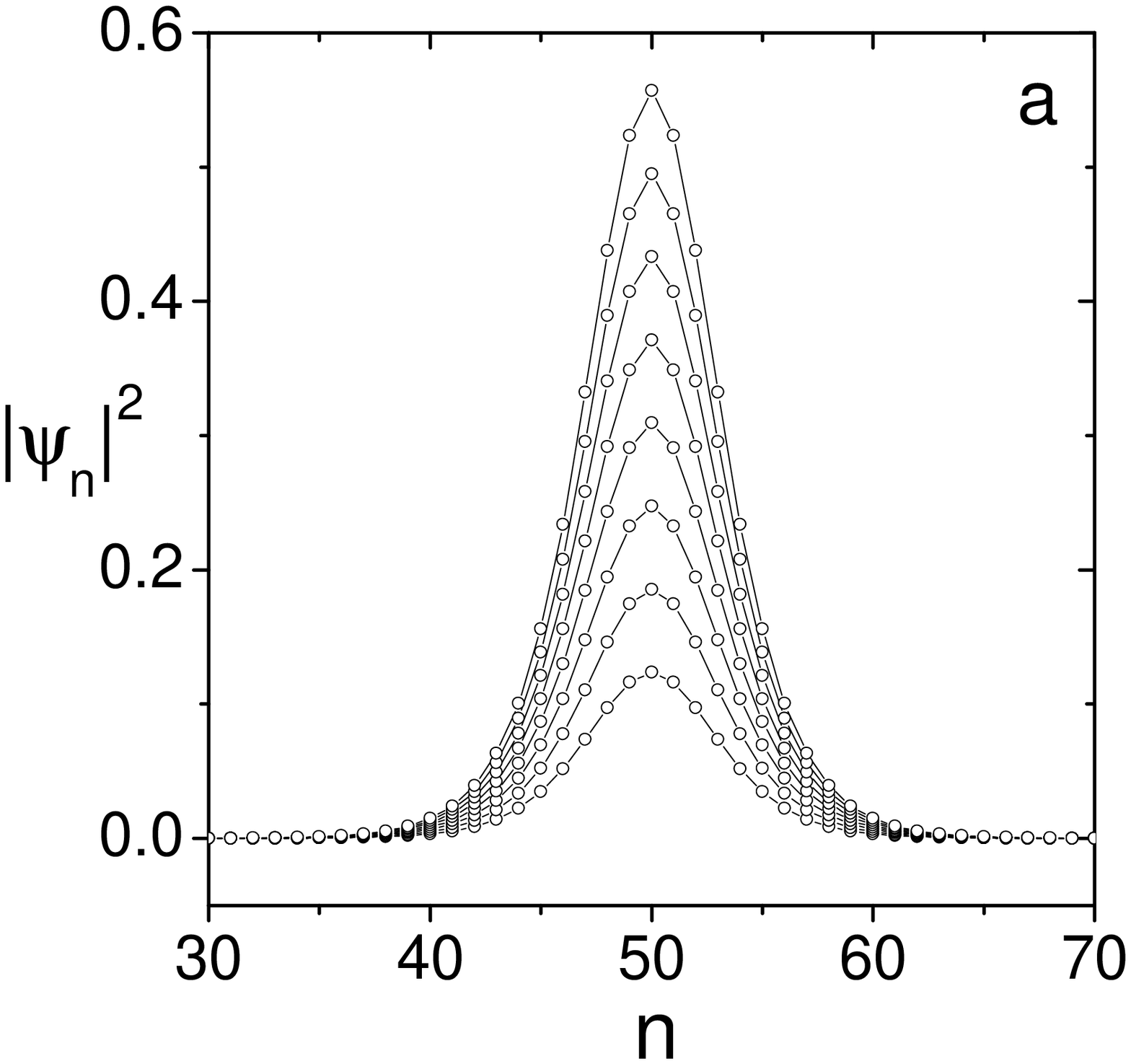}
\includegraphics[width=4.4cm,height=4.4cm,angle=0,clip]{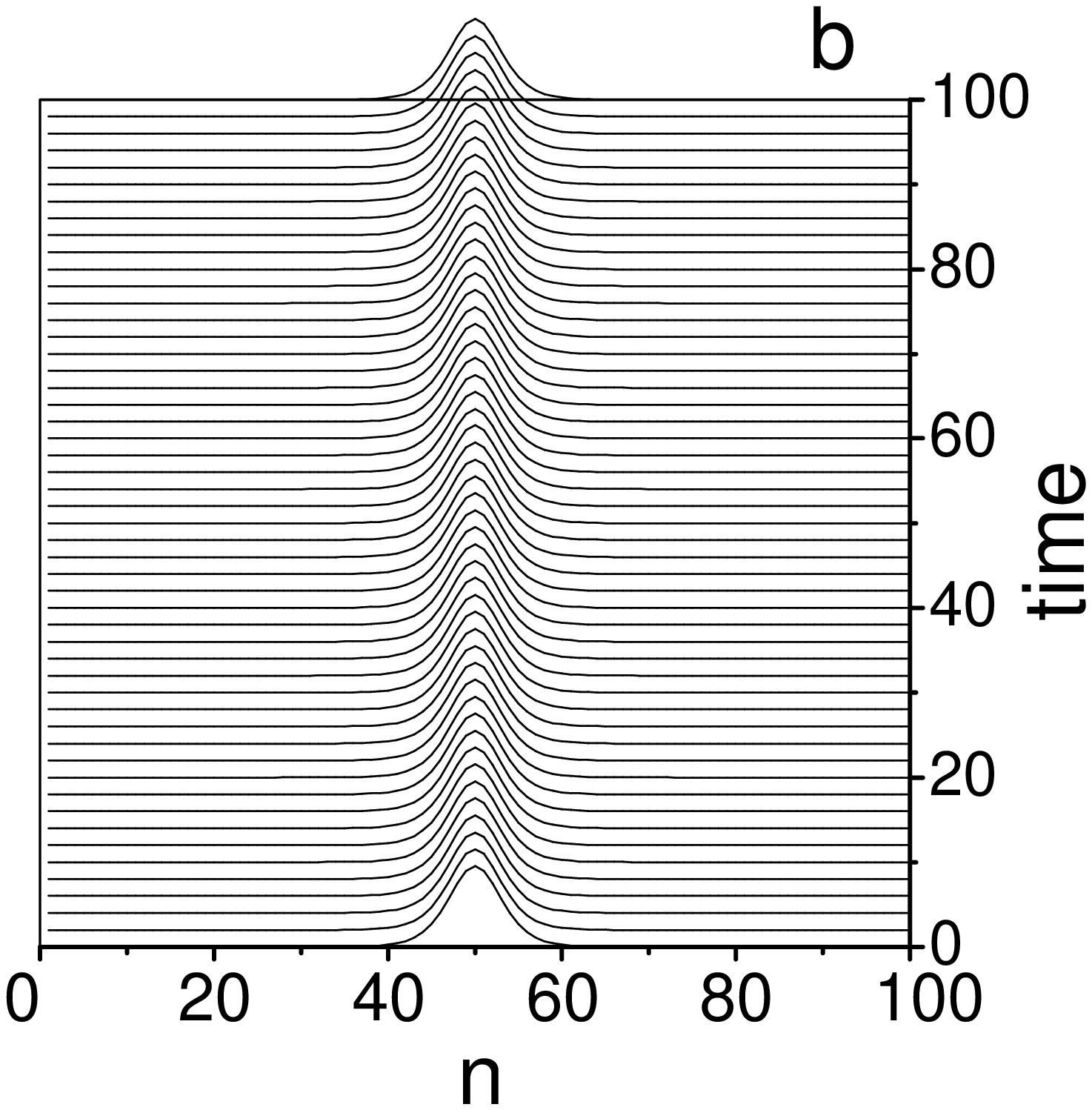}}
\centerline{
\includegraphics[width=4.2cm,height=4.2cm,angle=0,clip]{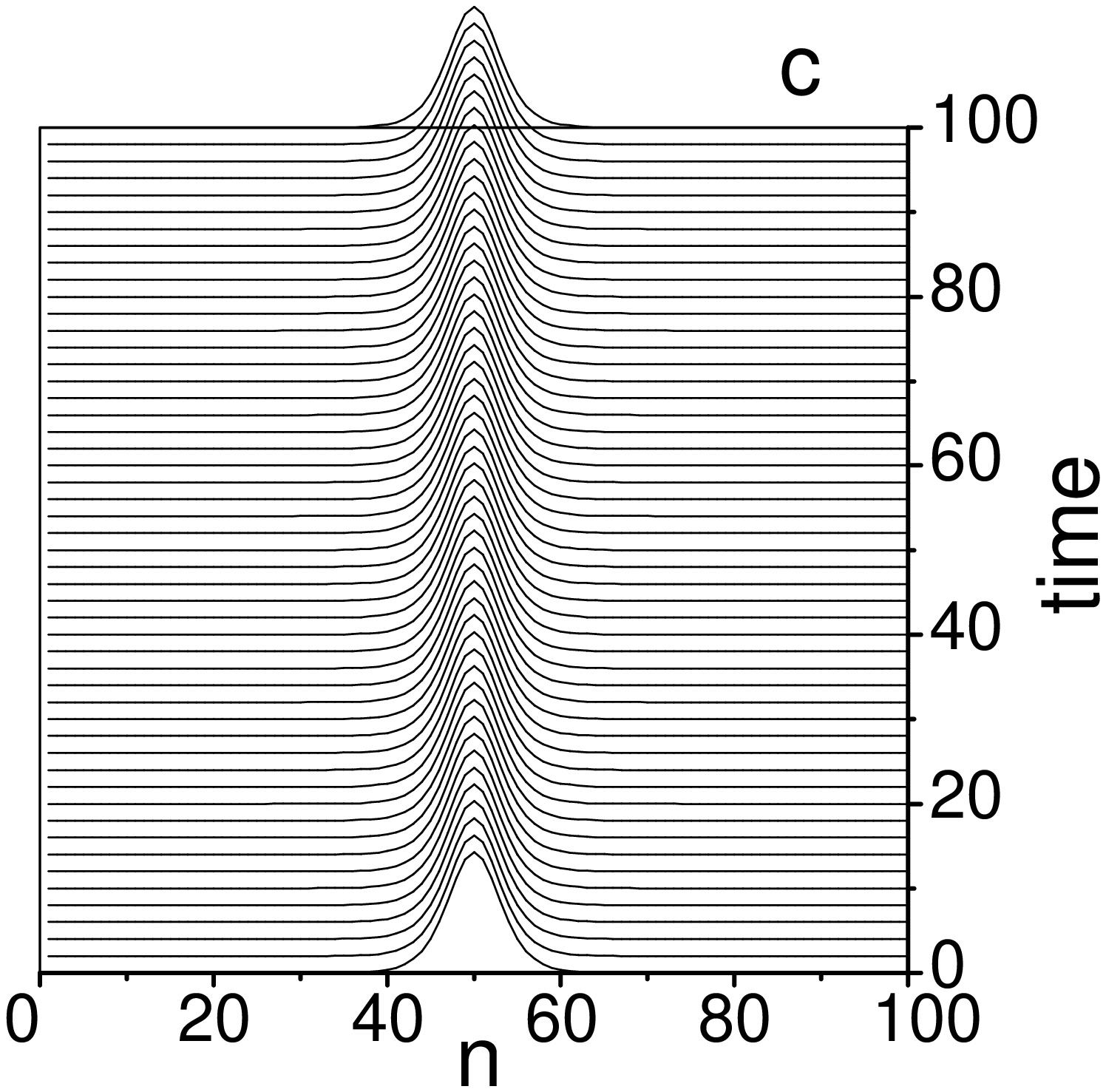}
\includegraphics[width=4.4cm,height=4.4cm,angle=0,clip]{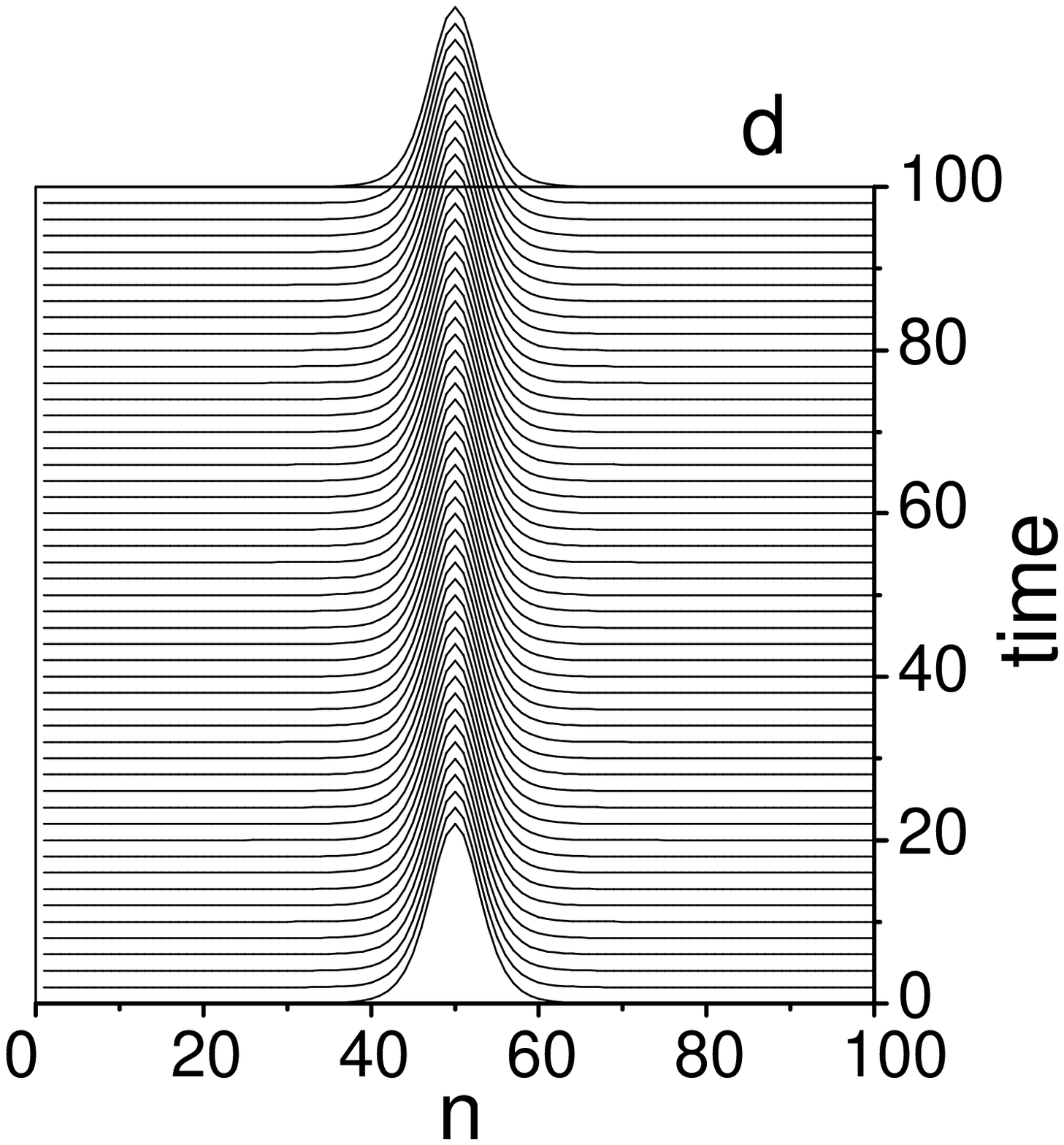}}
\caption{Panel (a). Stationary solutions of the GAL equation
($\nu=2$) for different values of $p$ ranging from $p=1$ (lower
curve) to $p=8$ (upper curve). Other parameters are $\beta=0.25$,
$m=1$. Panels (b), (c) and (d) show the time evolution of the
solutions corresponding to the cases $p=1,2,3$,
respectively. } \label{fig1}
\end{figure}

In the following we focus on the AL limit of Eq.
(\ref{Gen_Salerno_1}) and investigate the existence of localized
stationary solutions, discrete breathers and moving excitations,
by means of direct numerical integrations. In Fig. \ref{fig1}(a)
we show the profiles of the localized states corresponding to the
type III solution (soliton limit) for different values of $p$ and
for the same parameter $\beta$ [notice that this parameter fixes
the norm (power)]. We see that as $p$ is increased the amplitude
(and the norm) of the solution increases as a consequence of the
higher order nonlinearities. In panels (b), (c), and (d) of this
figure we also show the time evolution of the stationary solutions
for the cases $p=1,2,3$, respectively (similar results are
obtained for higher values of p).  Notice that the states are
quite stable under time evolution, thus confirming the existence
of stable localized solutions of the generalized Ablowitz-Ladik
(GAL) equation with higher order nonlinearity.

Next we concentrate on moving solutions and on discrete breathers.
To this end, we recall that for the case $p=0$ (i.e., the usual AL
equation) exact moving solutions of the traveling waves type are
well known \cite{Alan}. One can show that an extension of these
$p=0$ moving solutions to the case $p>0$ is not possible if one
assumes a traveling wave ansatz.
The existence of a zero PN barrier, however, strongly suggests the
existence of moving solutions for all values of $p$. In the
following we investigate this aspect by means of a direct numerical
experiment.  In this context, let us consider initial conditions
which are a linear superposition of two exact stationary solutions
of the form
\begin{eqnarray}
\psi(n,t)&=&A e^{-i\omega t} \left[ \dn[\beta (n-\frac N2+X_0)]+
\right.\nonumber \\ && \left. \dn[\beta (n-\frac N2-X_0)]
e^{-i\delta}\right], \label{inicond}
\end{eqnarray}
with $A, \omega$ given by our previous formulas. By properly
choosing the initial distance $2X_0$ between the centers of the
humps (in order to have a weak overlap) we can  bring them in
interaction and at the same time have a good initial condition
which is very close to an exact solution. As for the usual NLS
solitons, the interaction between the humps depends on their
distance and on their phase difference $\delta$. In particular, we
have that the two humps attract each other if they are in phase
($\delta=0$) and repel if they are out of phase ($\delta=\pi$).

\begin{figure}[t]\centerline{
\includegraphics[width=\columnwidth,angle=0,viewport=40 260 530 805,clip]{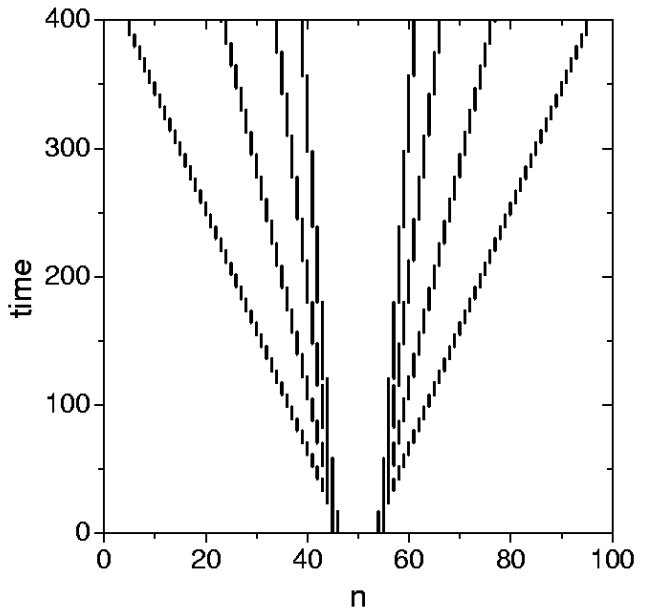}}
\caption{Time evolution of the position of the maxima of a two-hump
solution of the GAL equation with $p=1$, $m=1$, originated from
the initial condition in Eq. (\ref{inicond}) with the distance
between the two humps increased in steps of $1$ from $X_0=7$
(lower slopes) to $X_0=10$ (higher slopes). Other parameters are
fixed as $\delta=\pi$, $\beta=1.25$.} \label{fig2}
\end{figure}

The existence of a zero PN barrier can then be checked by
increasing the initial distance between two out-of-phase humps and
observe if the humps are set in motion by the mutual repulsion.
Since by increasing the initial distance one considerably  reduces
the interaction between the humps (the interaction goes to zero
exponentially with the distance) one has  that for large
separations motion can exist only if the PN barrier is zero. In
Fig. \ref{fig2} we show the results of such a numerical experiment
by reporting the trajectories of the humps center (point of maximum
amplitude) obtained from the GAL equation with $p=1$ (AL with
cubic-quintic nonlinearity) for  different values of $X_0$ and
initial phase $\delta=\pi$. We see that for small initial
separations the two humps move in opposite directions with high
velocity while as we increase the initial separation the velocity
gets progressively smaller. Our numerical investigations seem to
indicate the absence of any critical threshold in the initial
separation above which motion is stopped, which being in agreement
with our analytical considerations about the absence of the PN
barrier.

\begin{figure}\centerline{
\includegraphics[width=4.2cm,height=4.2cm,angle=0,clip]{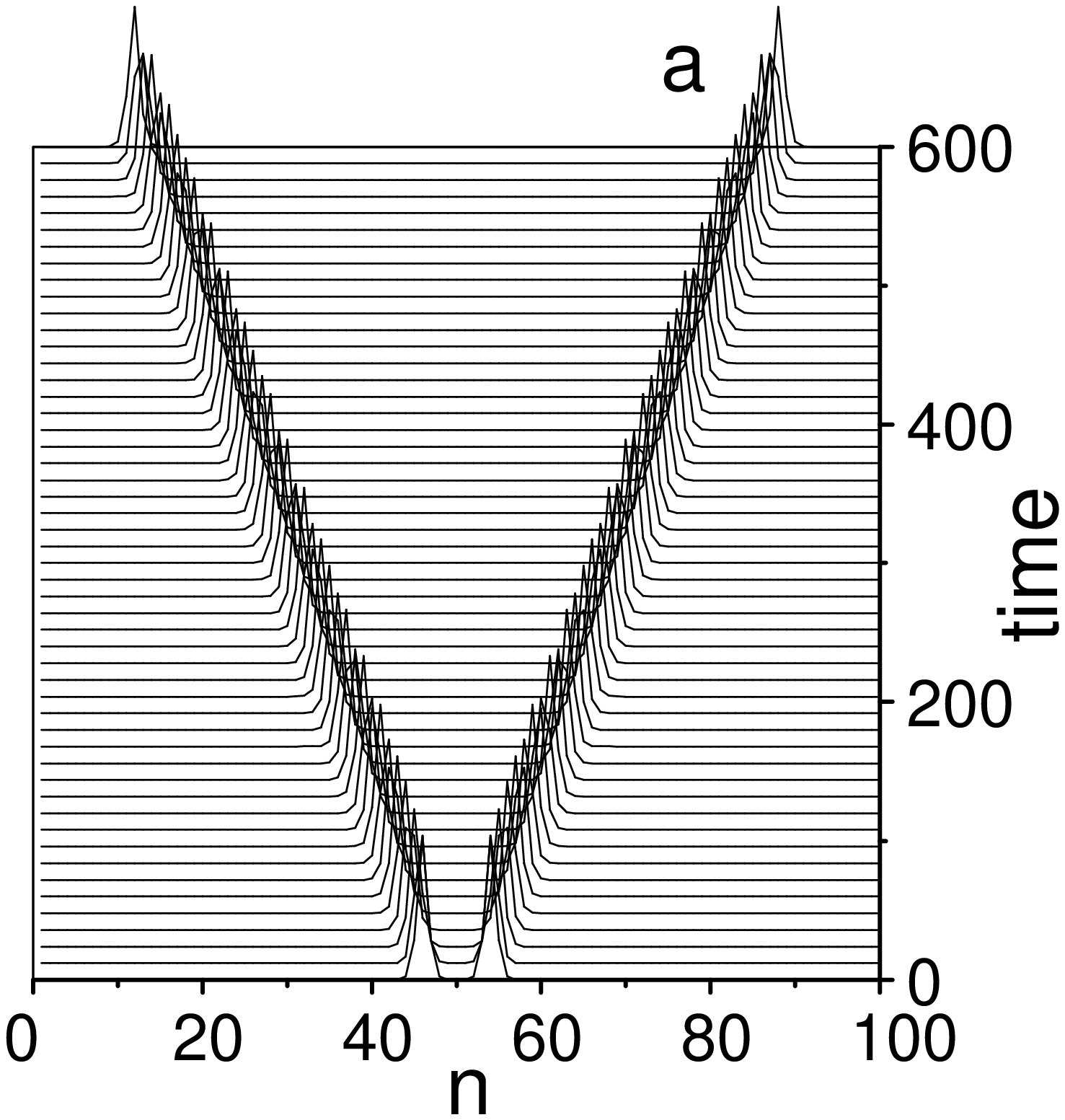}
\includegraphics[width=4.cm,height=4.cm,angle=0,clip]{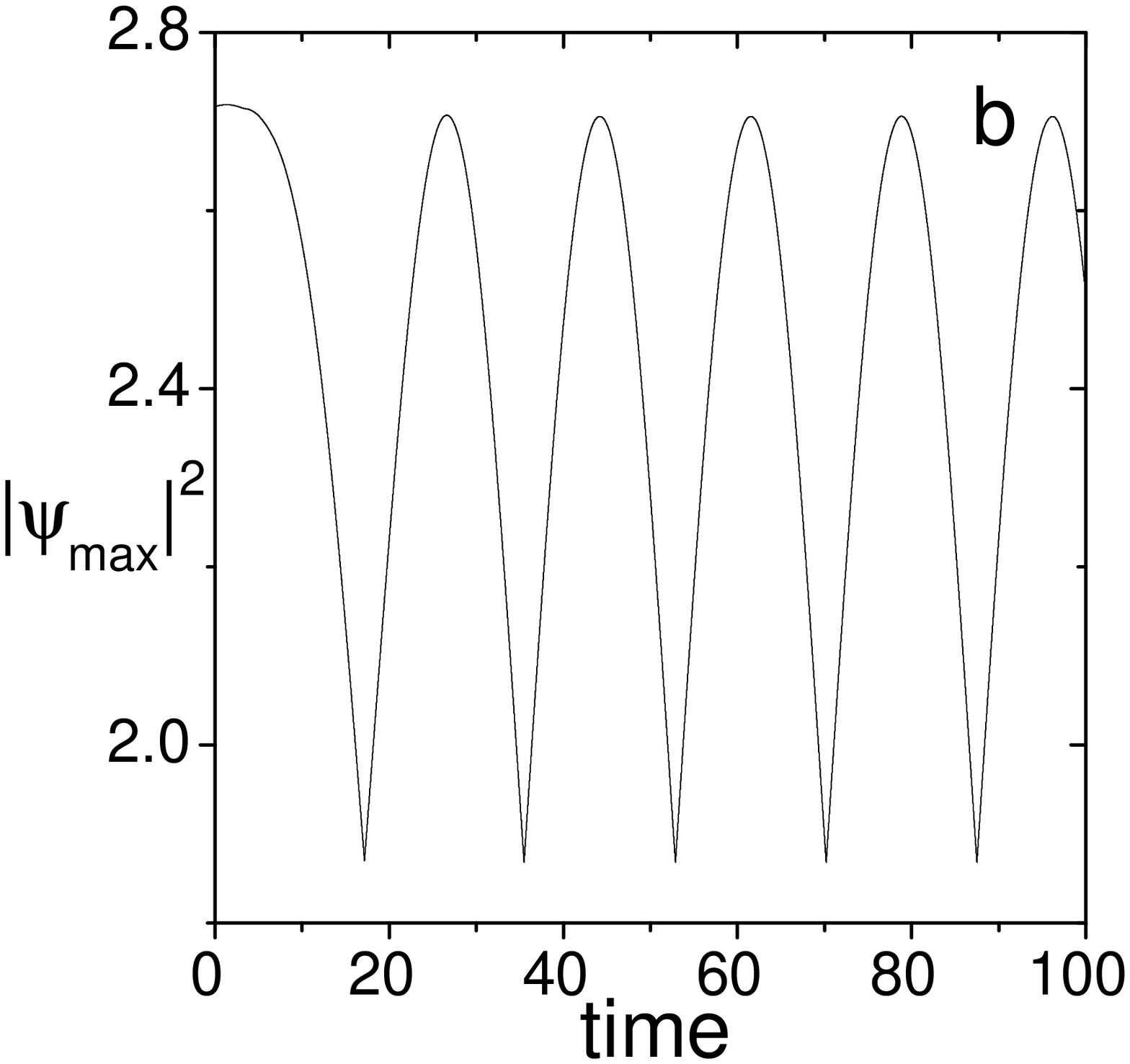}
}
\centerline{
\includegraphics[width=4.2cm,height=4.2cm,angle=0,clip]{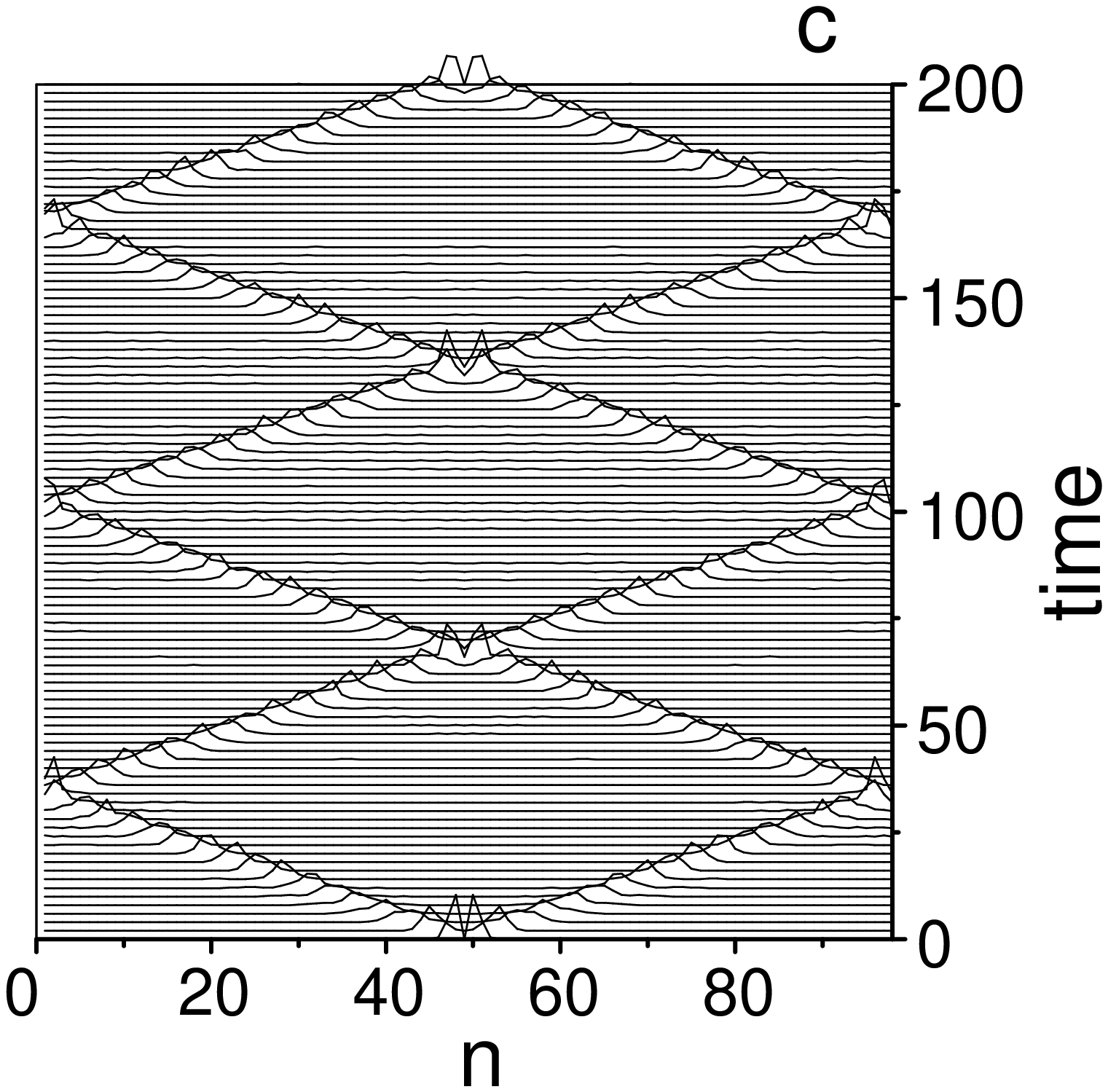}
\includegraphics[width=4.2cm,height=4.2cm,angle=0,clip]{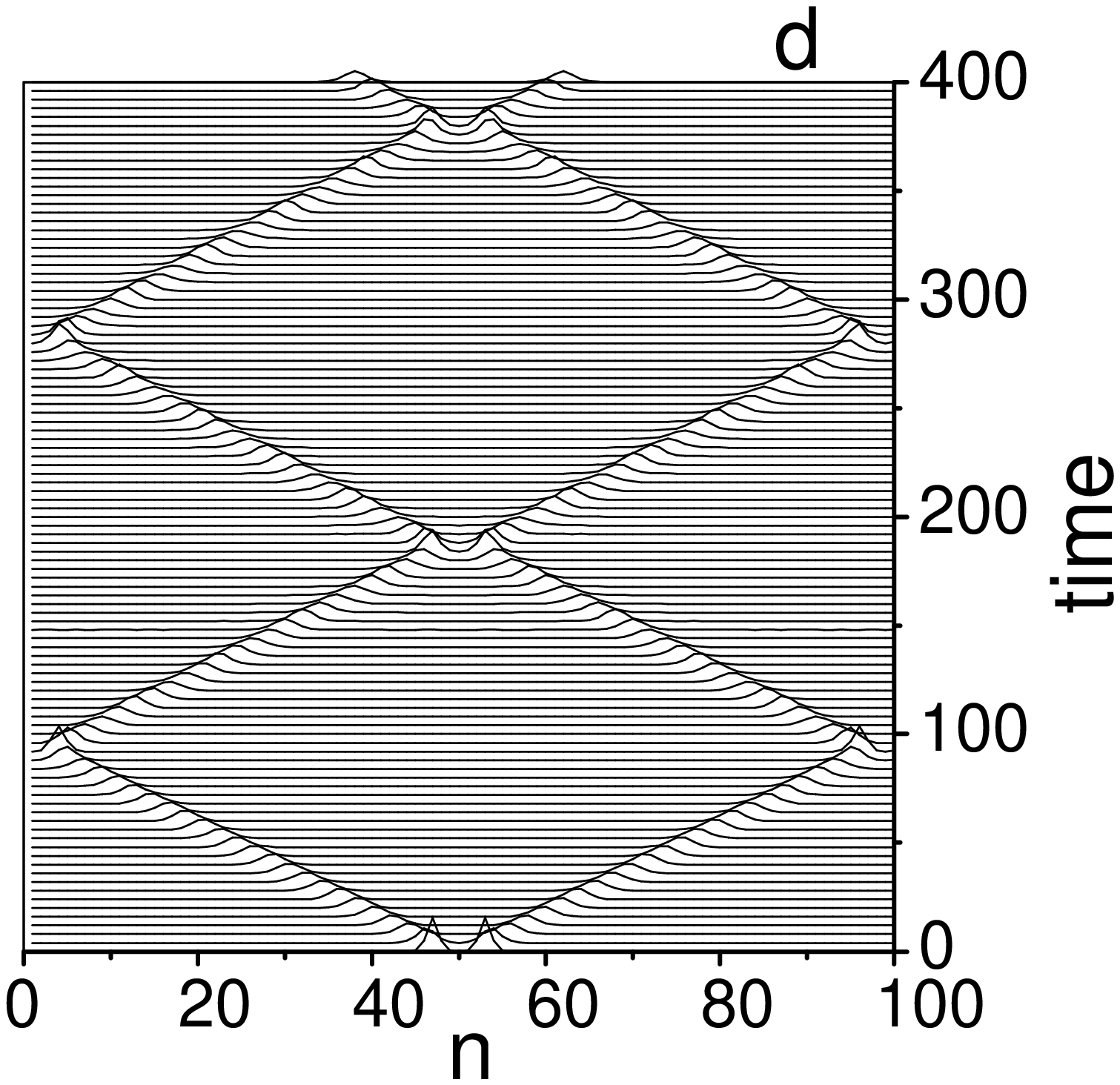}}
\caption{Panel (a). Time evolution of two solitons of the GAL
equation obtained for $p=1$, $m=1$ and initial condition taken as
in Eq. \ref{inicond} with $X_0=4$. Other parameters are fixed as
$\delta=\pi$, $\beta=1.25$. Panel (b). Time dependence of the
humps amplitude (modulo square of the maximum of the profiles)
depicted in the left panel. The minima correspond to the times at
which the amplitude moves to the next lattice site. Parameters are
the same as for panel (a). Panel (c). Time evolution of Eq.
(\ref{Gen_Salerno_1}) with $p=1$, $N=100$ and initial condition
taken as in Eq. (\ref{inicond}) with $X_0=1$, $m=1$, $\beta=1.25$
and $\delta=\pi$. Panel (d). Same as in panel (c) but for $p=2$
and $X_0=3$. } \label{fig3}
\end{figure}

This behavior is also seen from Fig. \ref{fig3}(a) in which the
time evolution of two stationary solutions of the GAL for $p=1$,
initially displaced by a distance larger than their rest widths,
is depicted. From this figure it is  clear that there is practically
no radiation generated during the motion, thus making the hump
dynamics very close to that of exact (traveling wave) solitons. In
panel (b) of this figure we depict the time dependence of the
amplitude during the hump motion of panel (a), from which we observe
that, except for the initial part where interaction dominates,  a
very regular pattern is generated. Notice that the amplitude
oscillation lobes correspond to the vertical segments visible in
the trajectories depicted in Fig. \ref{fig2}, the minima of the
lobes corresponding to the times at which the maximum of the profile
moves by one lattice site (i.e., to next vertical segment in the
trajectory plot). From this we infer that the reciprocal of the
period of the oscillation in the hump amplitude is just the hump
velocity in lattice site units.
It is remarkable that the absence of radiation in the hump
dynamics is observed also for very long times and can survive
multiple reflections. This is shown in panel (c) of Fig.
\ref{fig3} where the dynamics of two humps moving with  a higher
velocity (the velocity is increased by reducing the initial
separation) is depicted. Notice that the humps undergo several
collisions with almost no radiation generated. This behavior is a
direct consequence of the zero PN barrier and is reminiscent of
their soliton behavior. In panel (d) of Fig. \ref{fig3} we show
the same type of behavior for the case $p=2$. This result indicates
that moving localized states of the GAL equation may exist also for
higher values of $p$ (the stability of these solutions, however,
may become critical as $p$ is increased).
\begin{figure}\centerline{
\includegraphics[width=\columnwidth,viewport=60 500 530 750,clip]{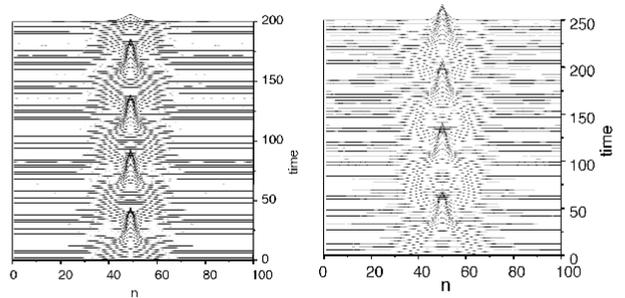}}
\caption{Left panel. Time evolution of Eq. (\ref{Gen_Salerno_1})
with $p=1$, $N=100$ and initial condition taken as in Eq.
(\ref{inicond}) with $X_0=5$, $m=1$, $\beta=0.15$ and $\delta=0$.
Right panel. Same as for the left panel but for $p=2$ and
$X_0=8$.} \label{fig4}
\end{figure}
The existence of a zero PN barrier makes it also possible to have
discrete breathers in the GAL equation. In order to show this we
repeat the same type of experiments considered above but with an
attractive instead of a repulsive interaction, i.e. we take the
initial humps to be in phase instead of out of phase.

In Fig. \ref{fig4} we depict the dynamics of two stationary solutions
of the GAL equation for $p=1$ and with a small value of $\beta$ so
that the initial profiles are very wide and have a large overlap.
We see that, due to the mutual attraction, the two localized
solutions undergo breathing oscillations with apparently no
radiation generated. Notice that the norm, which is fixed by the
parameter $\beta$, is below the instability threshold and the
oscillations continue forever. This solution represents therefore
an almost exact discrete breather of the GAL equation with higher
order nonlinearity. In the right panel of Fig. \ref{fig4} we show
a discrete breather for the case $p=2$, indicating that these
solutions may also exist with higher order nonlinearities. By
increasing the norm or by increasing $p$, however, we find that
an instability in the dynamics may develop at later times and the
solution may become unstable. A preliminary investigation
indicates the existence of a critical threshold (which
depends on $p$) below which stable stationary humps and discrete
breathers are stable. This stability threshold may be reminiscent
of the collapse threshold that exists in the continuous NLS with
higher order nonlinearities, for norms (powers) exceeding a critical
value. For most physical applications, however, only the lowest
higher order nonlinearities will be of interest (i.e., the cases
p=1,2). In these cases stable discrete soliton-like and discrete
breather solutions are found to be stable for a wide range of
parameters (which one should be able to check with linear stability
analysis).

\noindent
We remark that the absence of the PN barrier, the presence of
discrete breathers and the absence of emitted radiation during the
hump dynamics could indicate a possible complete integrability of
the GAL equation. Although this cannot be concluded without further
analysis, we remark that the vanishing of the PN barrier is a
necessary but not a sufficient condition for integrability. In this
context we remark that for the nonintegrable discrete $\phi^4$ models
\cite{oxtoby} and for discrete sine-Gordon chains \cite{zolo} it
is possible to have a zero PN barrier and radiationless moving
kinks for particular values of parameters. The GAL equation could
possibly be another example in which this phenomenon may occur.

\section{Conclusions}

We have introduced a class of discrete nonlinear
Schr\"odinger equations with arbitrarily high order nonlinearities
which include as particular cases the saturable discrete nonlinear
Schr\"odinger equation and the AL equation. We have obtained
three different types of exact solutions (both spatially periodic
in terms of Jacobi elliptic functions and their limiting
hyperbolic case) for these models as well as that of a higher
order generalization of the saturable discrete nonlinear
Schr\"odinger equation and the Ablowitz-Ladik equation. We then 
studied the Peierls-Nabarro
barrier \cite{PN, peyrard} for these solutions in various models
and found that it is zero indicating that the soliton-like
solutions move without experiencing any effect of the underlying
discreteness, which is quite remarkable. We also studied the
stability of these solutions under small perturbation and found
that they are robust as well as stable. Finally, we investigated
the collision of two hump solutions (both in-phase and
out-of-phase cases) and found that they collide and move without
any radiation. The out-of phase case indicates the formation of
discrete breathers. These results are strongly suggestive of the
integrability of the models introduced here, although we did not
attempt to prove this rigorously.  Our solutions and related
properties are likely to be useful in many physical contexts
including optical waveguides \cite{OP}, Bose-Einstein condensates
in optical lattices \cite{ST, ABDKS}, and  nonlinear optics in the
context of photonic crystals \cite{photonic}.

\section{Acknowledgments}
AK and MRS acknowledge the hospitality of the Center for Nonlinear
Studies and the Theoretical Division at Los Alamos.  MS wishes to
acknowledge partial financial support from a MURST-PRIN-2005
Initiative and the Department of Physics of The Technical
University of Denmark, Lyngby, for the hospitality. This work was
supported in part by the U.S. Department of Energy.

\end{document}